\def\paperID{0000}
\def\confName{ICCV}
\def\confYear{2025}

\def\paperTitle{Paper Title}

\def\authorBlock{
    Aiden Ochoa \qquad
    Xinyuan Xu\footnotemark[1] \qquad
    Xing Wang\thanks{Corresponding Author} \\
    Penn State University, Nuclear Engineering \\
    {\tt\small \{aio5165, xkx5062, xvw5285\}@psu.edu}
}

\newif\ifreview 
\newif\ifarxiv 
\newif\ifcamera \newcommand{\cameraready}{\cameratrue}
\newif\ifrebuttal 

\cameraready

\pdfoutput=1
\documentclass[10pt,twocolumn,letterpaper]{article}
\ifreview \usepackage[review]{cvpr} \fi
\ifarxiv \usepackage[pagenumbers]{cvpr} \fi
\ifrebuttal \usepackage[rebuttal]{cvpr} \fi
\ifcamera \usepackage[pagenumbers]{cvpr} \fi

\usepackage{graphicx}	
\usepackage{amsmath}	
\usepackage{amssymb}	
\usepackage{booktabs}
\usepackage{times}
\usepackage{microtype}
\usepackage{epsfig}
\usepackage{caption}
\usepackage{float}
\usepackage{placeins}
\usepackage{color, colortbl}
\usepackage{stfloats}
\usepackage{enumitem}
\usepackage{tabularx}
\usepackage{xstring}
\usepackage{multirow}
\usepackage{xspace}
\usepackage{url}
\usepackage{subcaption}
\usepackage{xcolor}
\usepackage[hang,flushmargin]{footmisc}

\ifcamera \usepackage[accsupp]{axessibility} \fi

% ...

\ifarxiv  \fi

\newcommand{\R}[1]{{%
    \textbf{%
        \ifstrequal{#1}{1}{\textcolor{red}{R#1}}{%
        \ifstrequal{#1}{2}{\textcolor{blue}{R#1}}{%
        \ifstrequal{#1}{3}{\textcolor{magenta}{R#1}}{%
        \ifstrequal{#1}{4}{\textcolor{teal}{R#1}}{%
                           \textcolor{cyan}{R#1}%
        }}}}%
    }%
}}

\usepackage{xr-hyper}

\makeatletter
\newcommand*{\addFileDependency}[1]{
  \typeout{(#1)}
  \@addtofilelist{#1}
  \IfFileExists{#1}{}{\typeout{No file #1.}}
}

\makeatother

\definecolor{cvprblue}{rgb}{0.21,0.49,0.74}
\usepackage[pagebackref,breaklinks,colorlinks,allcolors=cvprblue]{hyperref}
\usepackage[capitalize]{cleveref}
\crefname{section}{Sec.}{Secs.}
\crefname{table}{Table}{Tables}
\crefname{figure}{Fig.}{Figs.}

\ifarxiv \crefname{appendix}{App.}{Apps.}
\else \crefname{appendix}{Suppl.}{Suppls.} \fi

\usepackage[mathscr]{euscript}
\usepackage{hyperref}
\usepackage[accsupp]{axessibility}
\newcommand{\citesubfig}[2]{\hyperref[#1]{\ref*{#1}#2}}

\def\authorBlock{
	Aiden Ochoa \qquad
	Xinyuan Xu \qquad
	Xing Wang\thanks{Corresponding Author} \\
	Ken and Mary Alice Lindquist Department of Nuclear Engineering, Penn State University \\
	{\tt\small \{aio5165, xkx5062, xvw5285\}@psu.edu}
}
\def\paperTitle{Improving U-Net Confidence on TEM Image Data with L2-Regularization, Transfer Learning, and Deep Fine-Tuning}

\begin{document}
	\title{\paperTitle}
	\author{\authorBlock}
	\maketitle
	
	\begin{abstract}
With ever-increasing data volumes, it is essential to develop automated approaches for identifying nanoscale defects in transmission electron microscopy (TEM) images. However, compared to features in conventional photographs, nanoscale defects in TEM images exhibit far greater variation due to the complex contrast mechanisms and intricate defect structures. These challenges often result in much less labelled data and higher rates of annotation errors, posing significant obstacles to improving machine learning model performance for TEM image analysis. To address these limitations, we examined transferring learning by leveraging large, pre-trained models used for natural images. 

We demonstrated that by using the pre-trained encoder and L2-regularization, semantically complex features are ignored in favor of simpler, more reliable cues, substantially improving the model performance. However, this improvement cannot be captured by conventional evaluation metrics such as F1-score, which can be skewed by human annotation errors treated as “ground truth”. Instead, we introduced novel evaluation metrics that are independent of the annotation accuracy. Using grain boundary detection in UO$_2$ as a case study, we found that our approach led to a 64\% increase in the total number of grains detected, which acts a robust and holistic measure of model performance on the TEM dataset used in this work. Finally, we showed that model self-confidence is only achieved through transfer learning and fine-tuning of very deep layers.
\end{abstract}
	\section{Introduction}
\label{sec:intro}

Nanoscale defects, such as grain boundaries, precipitates, and dislocations, play a critical role in controlling the properties and functionality of solid-state materials. Transmission electron microscopy (TEM) has become an irreplaceable tool for investigating these defects, owing to its ultrahigh spatial resolution (sub-Angstrom) \cite{Williams2009}. Furthermore, recent advances in faster electron detection and data processing have enabled a big-data approach to characterization techniques such as in-situ TEM and 4D-scanning transmission electron microscopy (4D-STEM) \cite{Ophus2019}. Terabytes of data can be created in a single-hour session during 4D-STEM or in-situ TEM experiments \cite{Jacobs2022}. However, extracting meaningful insights from the TEM images has quickly become an enormous bottleneck, since the traditional method of manual image analysis is time-consuming, subject to human bias, and cannot scale with the growing data volume  \cite{Shen2021_1}. Therefore, developing high-quality automated approaches for TEM image analysis is of paramount importance.

Since the discovery of convolutional neural networks (CNN), machine learning (ML) models have been able to outperform not only traditional computer vision techniques, but even human abilities on certain image analysis tasks \cite{McKinney2020}. In the context of TEM, binary segmentation utilizing models from the U-Net family enables pixel-level classification of defect structures, which is especially important for identifying continuous defects like grain boundaries and phase interfaces \cite{Shen2021_2}. However, comparing applications to more typical datasets from other sciences, the performance of CNN models for TEM image analysis remains inferior. For instance, most of the literature involving U-Net based models, including the seminal paper \cite{Ronneberger2015} and popular derivatives like U-Net++ \cite{Zhou2018}, is focused on medical imaging \cite{Siddique2021}. With these datasets, it is very common to see F1-scores in the range 0.85-0.95. Similar performance can also be seen in applications like forestry \cite{Wagner2019}, crack detection \cite{Liu2019}, satellite imaging \cite{Alsabhan2022}, and plant disease \cite{Wang2021}. TEM applications, however, usually report lower F1-scores in range 0.5–0.8 \cite{Alrfou2022}\cite{Wang2022}.

One key reason for this performance gap stems from the inherent complexity of TEM image analysis. Unlike optical images, contrast in TEM arises from multiple mechanisms, making feature identification highly sensitive to imaging conditions and sample characteristics. Consequently, TEM datasets differ from conventional ML datasets in two major ways. (1) Smaller dataset size. TEM datasets typically contain tens to hundreds of  labeled images due to the time-intensive nature of annotation. In contrast, datasets like ImageNet \cite{Deng2009} contain more than 14 million annotated images. (2) Greater annotation ambiguity. The complex contrast mechanisms and intricate defect structures often result in large variations in how these defects appear in TEM images. This makes it challenging to annotate all defects within the images, leading  to considerable uncertainty and inconsistency in human-labeled data, which is nonetheless treated as "ground truth" during ML model training \cite{Bruno2023}.

\begin{figure*}[!t]
    \centering
    \includegraphics[width=0.75\linewidth]{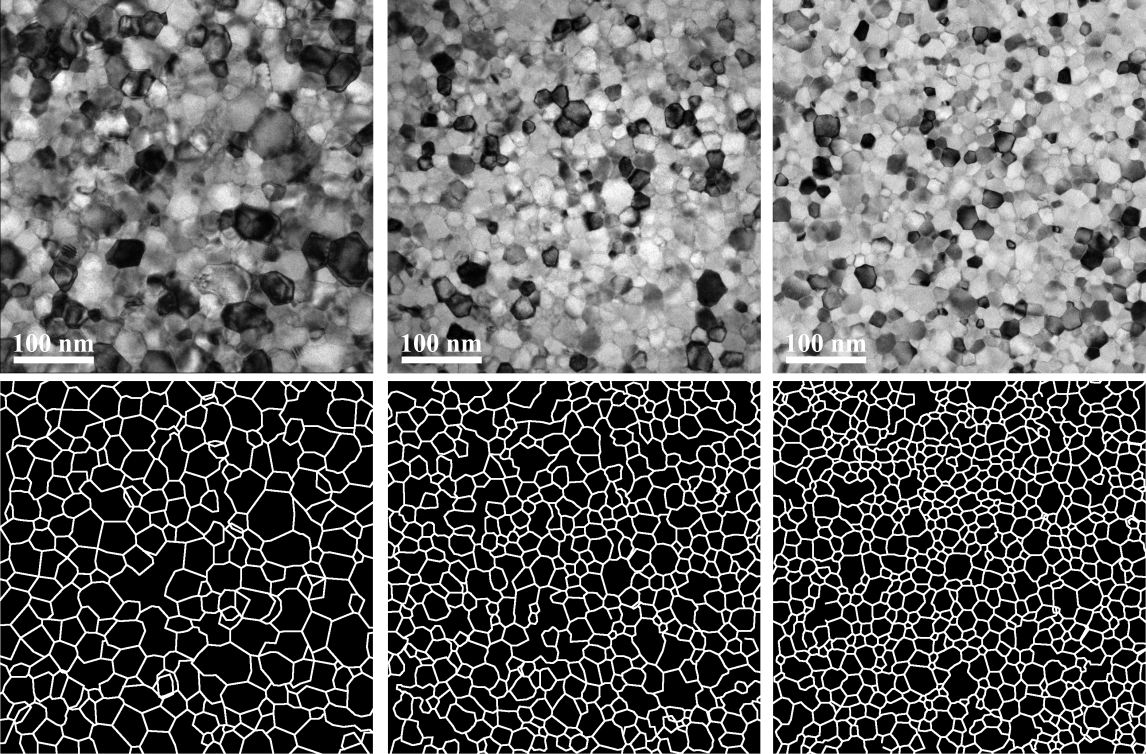}
    \caption{Representative TEM images of nanocrystalline UO2 (top) with corresponding annotated grain boundaries (bottom) from the dataset used in this work}
    \label{fig:data-comparison}
\end{figure*}

To address these challenges, we explored the use of transfer learning by leveraging advanced ML models pre-trained on ImageNet. In particular,  we investigated promising combinations of a pre-trained EfficientNet encoder with a U-Net++ decoder, which helped achieve the best performance in 2023 and 2024 for edge detection in regular optical image datasets \cite{Zhou2023}\cite{Zhou2024}. We also introduced two novel metrics, prediction certainty and prediction abundance, describing the ability of a model to make predictions with high class probabilities. Together, they define a model’s self-confidence and are independent of ground truth accuracy. We demonstrated that U-Net performance can be substantially improved using a pre-trained encoder, fine-tuning of deep layers, and L2-regularization to control overfitting. Although F1-scores remained limited due to ground truth flaws and uncertainty, improved model self-confidence led to a 64\% increase in the number of detected grains for the dataset tested, even surpassing human abilities and  significantly enhancing the reliability of defect statistics.
 
As a practical example, we applied our workflow to a TEM image dataset of nanocrystalline UO$_2$ samples used to study grain growth as function of temperature and heavy ion irradiation dose \cite{Yu2022}. UO$_2$ is the primary fuel used in current nuclear reactors, and many of its key material properties, such as thermal conductivity, fission gas retention, and fracture toughness, are governed by its grain size. Therefore, it is critical to establish reliable correlations between grain size of UO2 and its irradiation conditions for predicting the nuclear fuel performance. The accuracy of said correlations principally depends on the quality of grain statistics, which in turn requires large and representative datasets. As such, hundreds of TEM images like those in Fig. \ref{fig:data-comparison} were collected at various dose levels and temperatures. Processing these images by hand is infeasible, so ML models are needed to automate segmentation of grain boundaries. To broaden the scope of this study, our methodology was also applied to additional defect types, as discussed in Sec. B of supplementary materials. 
\begin{figure*}[!t]
    \centering
    \includegraphics[width=0.9\linewidth]{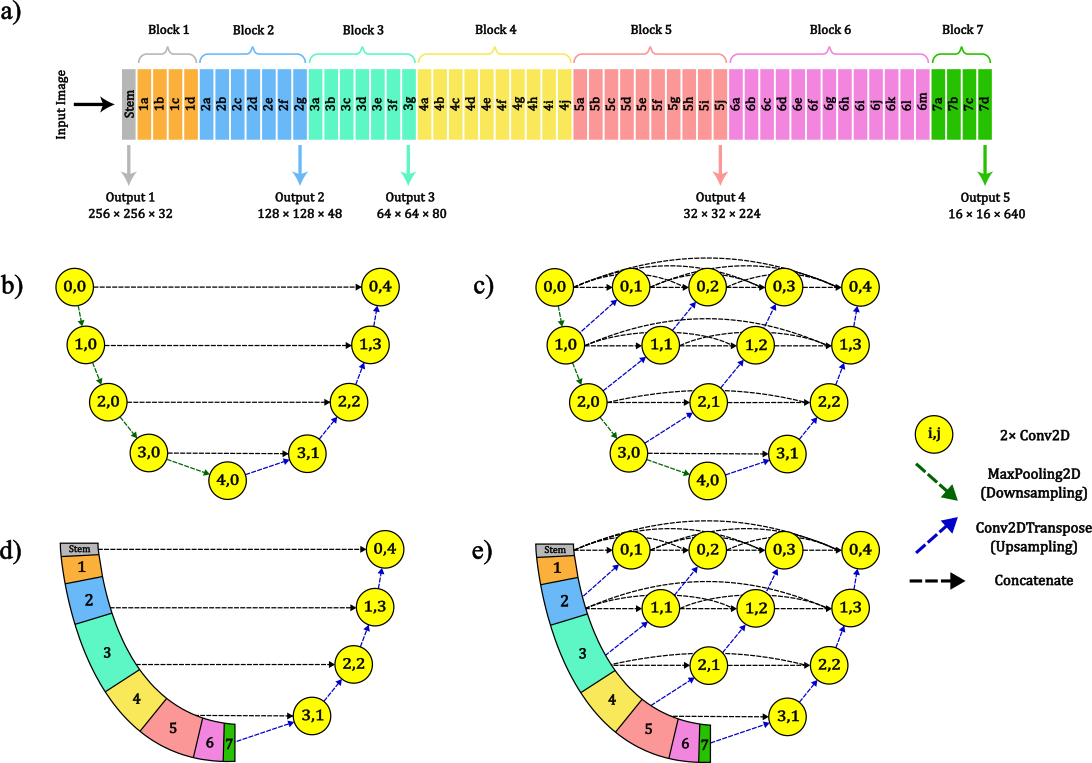}
    \caption{Model architectures evaluated. a) EfficientNetB7 (64M parameters) with 56 convolution layers. b) U-Net (31M) baseline architecture. c) U-Net++ (36M) with intermediate convolution blocks and dense connections. d) U-Net with an EfficientNetB7 backbone (72M). e) U-Net++ with an EfficientNetB7 backbone (76M)}
    \label{fig:architectures}
\end{figure*}

\section{Methodology}
\label{sec:methodology}

\subsection{Grain Boundary Dataset}

The dataset used in this work consists of bright-field (BF) TEM images of nanocrystalline UO$_2$ samples. Details about fabrication, in-situ irradiation, and TEM imaging can be found in \cite{Yu2022}\cite{Xu2024}. The training dataset was created by selecting 14 images from a variety of experimental conditions and resizing them from 4096$\times$4096 to 1024$\times$1024. The grain boundaries were then labeled using Inkscape with a thickness of 5 pixels. Pixels belonging to a grain boundary were given a value of 1, while pixels belonging to the grain interior were given a value of 0. A one-to-one mapping is ultimately constructed between TEM and annotated images. Examples can be seen in Fig. \ref{fig:data-comparison}.

Small dataset size can lead to significant variation in model performance when different validation and test hold-out sets are used. To improve their representation of the training set, each image was quartered into four 512$\times$512 images, and 5-fold cross-validation was used. This divides the dataset into five non-overlapping validation sets of equal size, training a model for each one with the same hyperparameters. Statistics can then be calculated for performance metrics, providing a more robust estimate. This also means a test set is not necessary, further improving validation set representation. To improve model performance, training images were augmented eightfold using 90° rotations and flipping over horizontal and vertical axes. Each model thus used a different combination of 360 training images and 11 validation images. 

\subsection{Model Architectures and Hyperparameters}

To improve U-Net performance, the encoder is replaced with EfficientNetB7, the largest model from the EfficientNet-v1 family \cite{Tan2019}, and the decoder has been replaced with U-Net++. Most promisingly, EfficientNetB7-U-Net++ has been used as the base network of the state-of-the-art edge detectors UAED \cite{Zhou2023} and MUGE \cite{Zhou2024}, which are applicable to grain boundary segmentation. Furthermore, EfficientNetB7 has been pre-trained on ImageNet \cite{Deng2009} to ease training and boost encoder performance for small datasets. The benefits of transfer and fine-tuning are evaluated in Sec. \ref{sec:encoder}. EfficientNetB7 and U-Net++ were also isolated to show their respective contributions, leading to the four architectures investigated in this study, as summarized in Fig. \ref{fig:confidence}: UNet (b), UNet++ (c), Eff-UNet (d), and Eff-UNet++ (e).  

The Adam optimizer was used with a global learning rate of 1e-4, as preliminary testing showed that this provided sufficient convergence speed, while having minimal loss fluctuation. It was also found that batch size minimally affected validation loss due to the already large image size (512$\times$512), so a batch size of 4 was used to improve batch representation while still offering frequent weight updates. For the yellow convolution blocks in Fig. \citesubfig{fig:architectures}{b-e}, each Conv2D layer learned $2^(i+6)$ filters, where $i=0,1,2,3,4$ refers to the encoder resolution level. ReLU activation was used for all layers except the output layer which used Sigmoid activation. To suppress overfitting, L2-regularization was used to penalize large weights in favor of a more distributed network. The total loss can be written symbolically as $$\mathscr{L}=\textrm{BCE}(\textrm{prediction},\textrm{annotation})+ \lambda||\mathbf{w}||^2,$$ where BCE is binary cross-entropy loss, commonly used in binary segmentation, $\mathbf{w}$ is a vector of model weights, and $\lambda$ is the L2-regularization strength. In preliminary testing, it was found $\lambda$ can have substantial effects on model output, specifically on model self-confidence. Given that the architectures in Fig. \ref{fig:architectures} have different sizes and connections, and thus very different $\mathbf{w}$, a coarse grid search was performed to determine the optimal $\lambda$. This grid search simultaneously provided the basis for model comparison. 
 
\subsection{Performance Metrics}

\begin{figure*}[!b]
    \centering
    \includegraphics[width=0.9\linewidth]{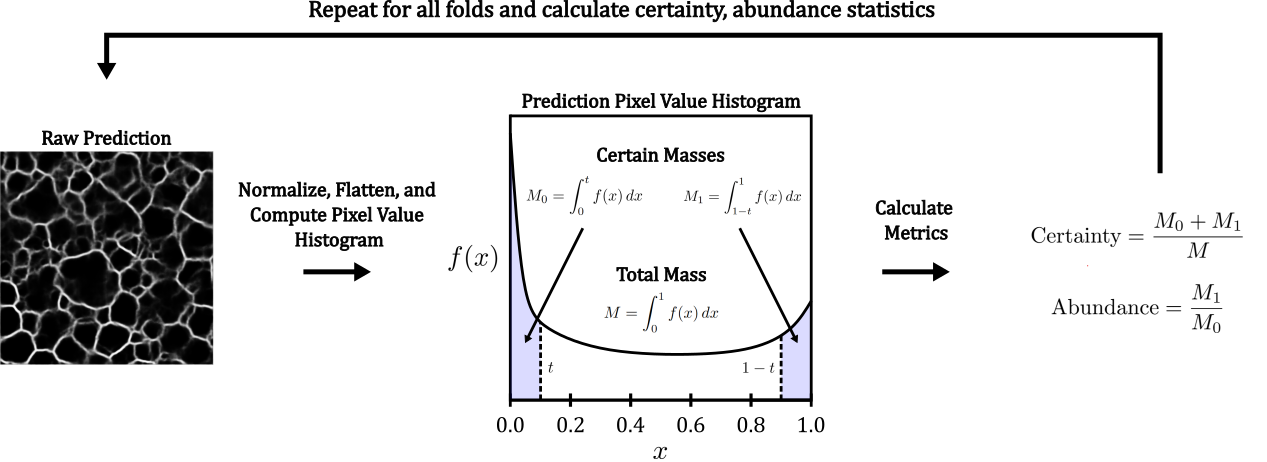}
    \caption{Proposed metrics, certainty and abundance, calculated from pixel value histograms of raw model predictions. The confidence threshold t=0.15 was used in this work}
    \label{fig:confidence}
\end{figure*}

To evaluate model performance, classification metrics, i.e., precision, recall, and F1-score, were used with a binarization threshold of 0.5. Two novel metrics, prediction certainty and abundance, were also used to evaluate model self-confidence, defined as the model’s ability to be certain of its own predictions regardless of classification performance. As shown in Fig. \ref{fig:confidence}, the proposed metrics operate on raw prediction image histograms, where certainty measures the ability of a model to predict high class probabilities, and abundance measures the proportion of said pixels that are positive predictions. A model with high certainty tends to avoid uncertain predictions where a pixel could belong to either a grain boundary or non-grain boundary with nearly equal probability. A model with high abundance tends to predict more grain boundaries overall, which is important for a class-imbalanced dataset. We describe model self-confidence comparatively as the possession of both certainty and abundance, which, importantly, do not rely on the existence of a ground truth. 

The final metric, detected defect count, was used to provide an unsupervised, holistic measure of model usefulness, as it operates on post-processed model predictions. For the grain boundary dataset, the post-processor Convex Hull Approximate Contour (CHAC) algorithm developed recently in previous work \cite{Xu2024} was used. CHAC uses traditional computer vision techniques to identify approximately closed and convex contours in model predictions as grains, based on thermodynamic principles for grain growth. As such, models that can produce segmentation maps leading to a greater number of grains identified by CHAC are more useful since grain size statistics will be more reliable and accurate. Detected defect count therefore implicitly depends on all ML metrics evaluated, as a model must have good classification performance and self-confidence to predict well-defined and crisp grain boundaries. Finally, because each TEM image, and thus each validation set, has a different number of grains, the total number of defects were tabulated across the entire dataset, made possible by $k$-fold cross-validation.

To examine model performance more comprehensively, additional supervised holistic metrics describing sample accuracy, including intersection-over-union and L1 distance metrics, were employed with the results discussed in Sec. A of supplementary materials.

\begin{figure*}[!t]
    \centering
    \includegraphics[width=0.99\linewidth]{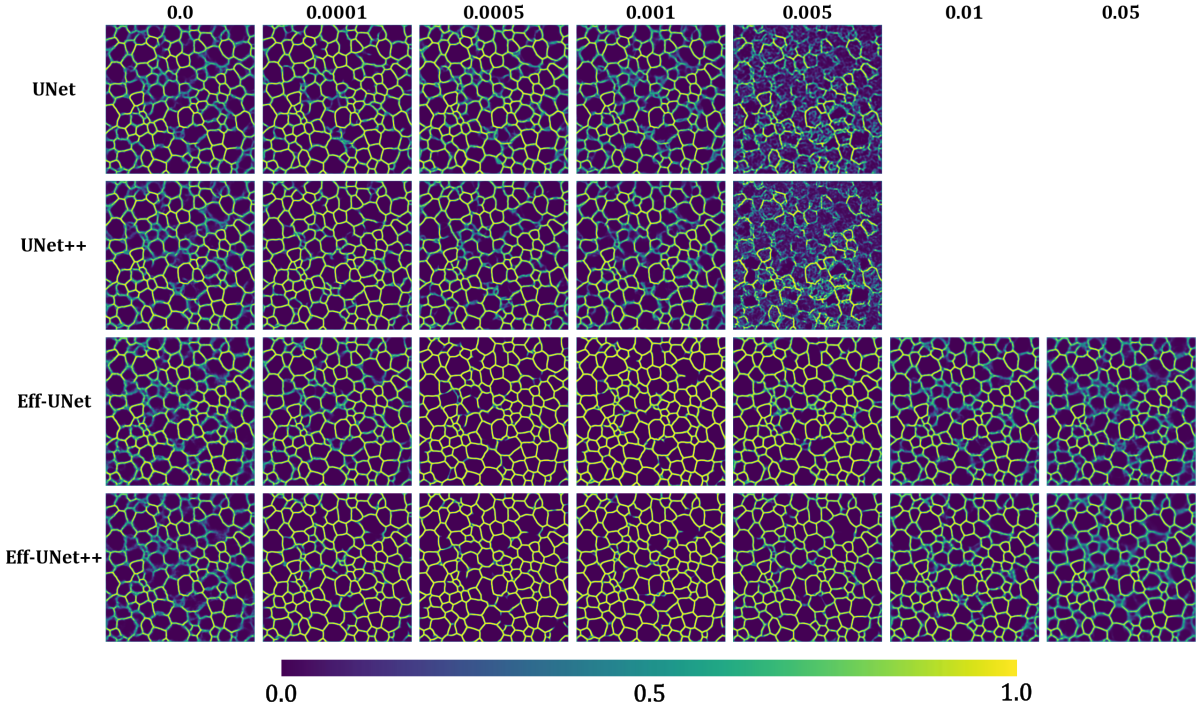}
    \caption{Comparison of raw validation predictions by U-Net architecture and L2-regularization strength $\lambda$ for a representative TEM input image}
    \label{fig:compare_pred}
\end{figure*}

	\section{Results and Discussion}
\label{sec:results}

\subsection{Visual Comparison of Predictions}

Initial insights into model performance can be gained by examining raw model predictions visually. A representative validation prediction is shown for all architectures and L2-regularization strengths in Fig. \ref{fig:compare_pred}. When no regularization was used, all architectures appeared to perform similarly. However, upon increasing $\lambda$, UNet and UNet++ responded quite differently than Eff-UNet and Eff-UNet++. The latter architectures exhibited greater tolerance to regularization, as over-regularization occurred at much higher $\lambda$ values. Furthermore, predictions made near  $\lambda$=1e-3 demonstrated significant quality improvements, where models clearly predict higher class probabilities, and thus have greater certainty. This has the primary benefit of reducing grain boundary noise in binarized predictions, as predicted probabilities are farther from the threshold (0.5), and are therefore less susceptible to random noise. Finally, unlike changing the encoder, there was no obvious difference between using U-Net and U-Net++ decoders. 

\subsection{Performance Metrics Analysis}

\begin{figure*}[!t]
    \centering
    \includegraphics[width=0.95\linewidth]{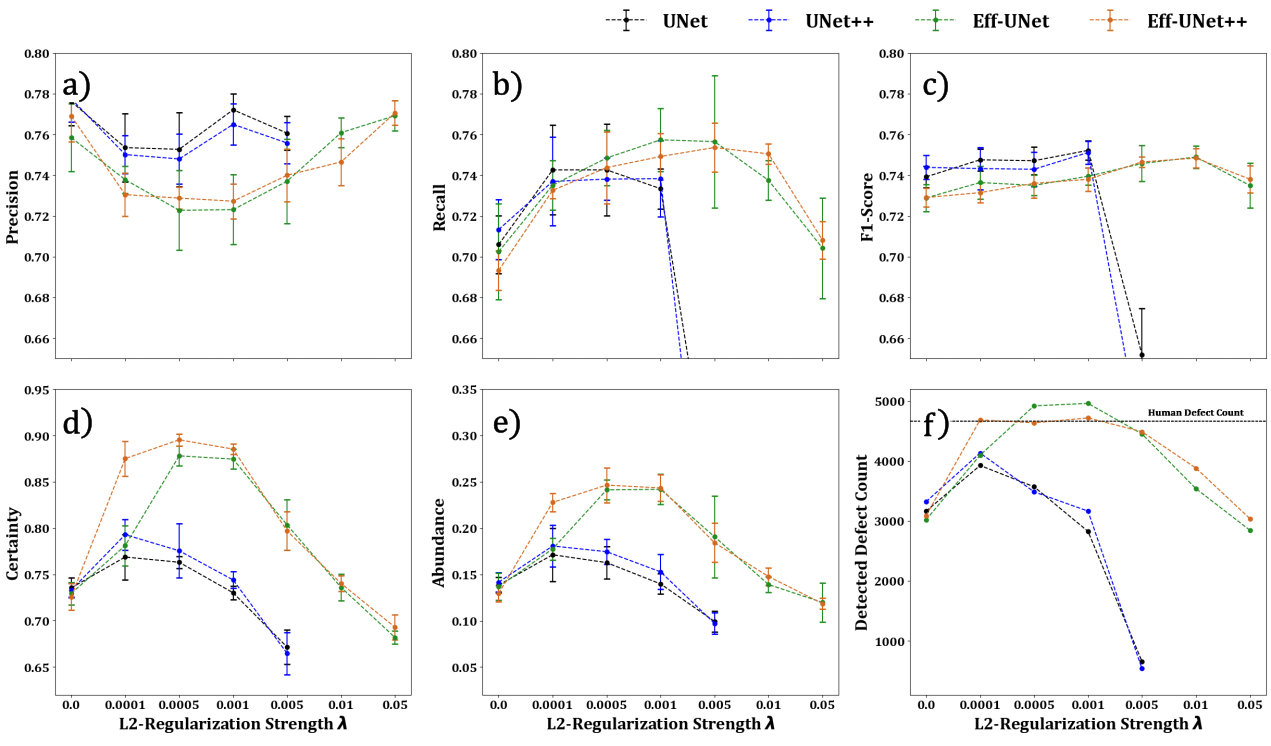}
    \caption{Metrics obtained from 5-fold cross-validation for $\lambda$ optimization. Figs. a-c are classification metrics precision, recall, and F1-score respectively. Figs. d-e are proposed metrics certainty, abundance, and detected defect count respectively. Error bars correspond to $\pm 1 \sigma$}
    \label{fig:l2-optimization}
\end{figure*}

Fig. \ref{fig:l2-optimization} summarizes the six metrics used to evaluate model architectures with L2-regularization strengths ($\lambda$). Like the visual results in Fig. \ref{fig:compare_pred}, architectures behave similarly when $\lambda$=0, other than Eff-UNet and Eff-UNet++ having marginally lower precision and recall. As $\lambda$ is increased, recall, certainty, abundance, and detected defect count reached their maximum at $\lambda$=1e-4 for UNet and UNet++, and between $\lambda$=1e-3 and $\lambda$=5e-4 for Eff-UNet and Eff-UNet++. Coincidentally, precision reached its minimum at these points. Furthermore, the following observations can be made: (1) F1-score showed little-to-no improvement despite considerable improvement to detected defect count, (2) there is a large divergence between U-Net and EfficientNet encoder architectures, with the latter showing much greater certainty, abundance, and detected defect counts in Figs. \citesubfig{fig:l2-optimization}{d-f}, and (3) given the width of the error bars, there is no practical benefit from using the U-Net++ decoder. 

Most significantly, by using EfficientNet and L2-regularization, detected defect count increased by 64\% over unregularized UNet, even outperforming humans by about 300 grains. This is clearly captured by improvements to certainty and abundance, but the conventional metrics in Figs. \citesubfig{fig:l2-optimization}{a-c} do not explicitly show this, other than a 7\% improvement to recall. This highlights the value of model self-confidence and holistic metrics like detected defect count. As discussed in the next section, the F1-score may not be entirely reliable for images that have higher ambiguity and the human annotation contains more errors. 

\subsection{Importance of L2-Regularization}\label{sec:l2_reg}

\begin{figure*}[!t]
    \centering
    \includegraphics[width=0.9\linewidth]{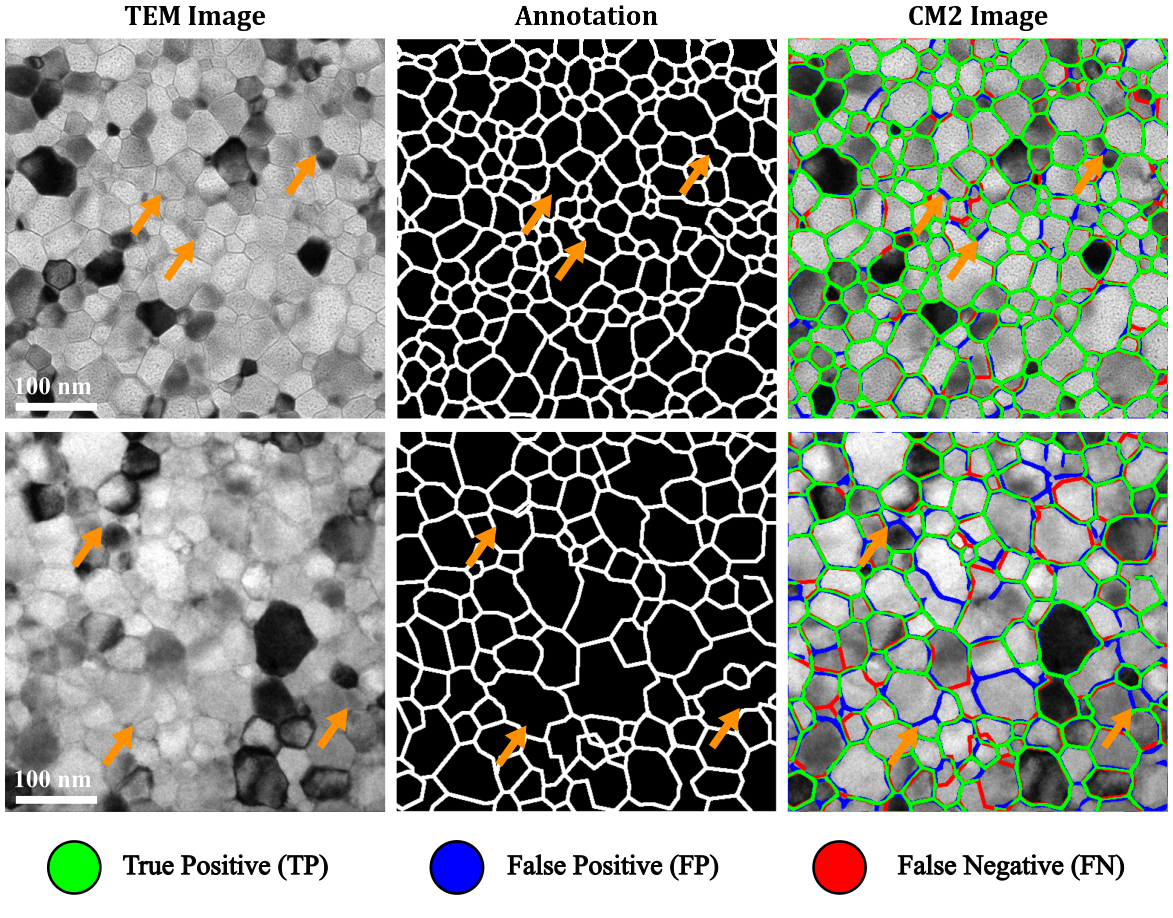}
    \caption{Model predictions with Confusion Matrix Color Mapping (CM2) and their corresponding TEM image and annotation. Orange arrows indicate examples of annotations error. Predictions come the validation sets of Eff-UNet models with $\lambda$=1e-3}
    \label{fig:cm2}
\end{figure*}

As shown in Fig. \citesubfig{fig:l2-optimization}{c}, the F1-score is limited to 0.75, regardless of architecture or regularization strength. To investigate this limitation, classification performance was visualized in Fig. \ref{fig:cm2} using a technique we named Confusion Matrix Color Mapping (CM2). Here, every pixel in a binarized prediction image is colored according to its classification outcome in the binary confusion matrix, i.e., green for true positive, blue for false positive, and red for false negative pixels. Importantly, orange arrows indicate examples of grain boundaries which were well-supported by the TEM image and were detected by the Eff-UNet model, but not present in the human annotation. In other words, the human annotator missed these grain boundaries. Annotation errors act as a significant limitation of classification performance where good model predictions are incorrectly penalized. 

To minimize loss, a model attempts to learn a set of operations that replicate the human annotated grain boundaries. However, when true grain boundaries are mistakenly missing in the annotation, the model cannot rationalize these inconsistencies, potentially damaging model self-confidence. Any attempt to exactly fit these annotation errors acts as a contradictory force to its typically well-supported operations. If annotation errors are interpreted by the model as high-level features with complex semantic content, the model could begin to overfit by trying to learn these features. The benefits of L2-regularization then become obvious: by constraining model complexity, it discourages fitting such annotation errors, and can instead help the model prioritize simpler, more reliable cues, thus promoting model self-uncertainty. 

The theory above is well supported by quantitative results in Fig. \ref{fig:l2-optimization}. Without regularization, all four architectures have relatively high precision and low recall, meaning they guess less but are usually correct, and have very low prediction certainty and abundance. This indicates that they are skeptical; models prefer to guess less grain boundaries, but those that are guessed tend to be more correct. However, applying optimal L2-regularization, the architectures shift toward low precision and high recall, where they now guess more often and are less concerned with correctness. Certainty and abundance also increase substantially, suggesting less influence from human annotation errors and more confidence with their predictions. Among all four architectures, Eff-UNet and Eff-UNet++ show the most significant improvement to prediction quality with detected defect count reaching the highest observed levels, likely because of the greater capacity and redundancy supplied by EfficientNetB7.
	\section{Encoder Fine-Tuning}
\label{sec:encoder}

\begin{figure*}[!t]
    \centering
    \includegraphics[width=0.95\linewidth]{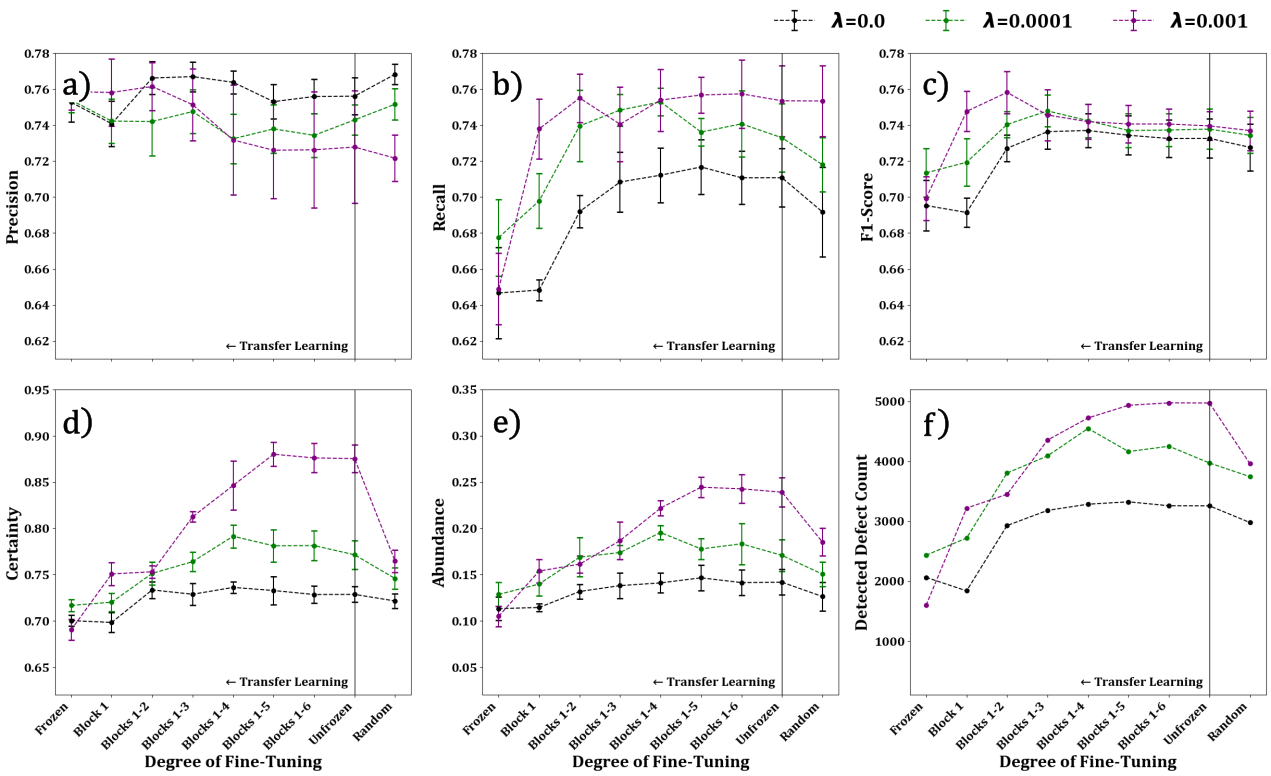}
    \caption{Metrics obtained from 5-fold cross-validation for fine-tuning optimization. The horizontal axis refers to which blocks of EfficientNetB7 (see Fig. \ref{fig:architectures}) are unfrozen. “Frozen” refers to no fine-tuning and “Random” refers to no pre-training. Errors bars correspond to $\pm 1 \sigma$}
    \label{fig:finetuning}
\end{figure*}

It has been shown that the performance of U-Net can be improved by using an EfficientNetB7 encoder pre-trained on ImageNet, with all weights unfrozen. However, it remains unclear whether the performance gains stem from the EfficientNet architecture itself, or from pre-training and fine-tuning. To address this question, model performance was studied for various degrees of encoder fine-tuning and L2-regularization. Starting from a fully frozen encoder with ImageNet weights, the blocks of EfficientNet were progressively unfrozen until all were trainable. To isolate the effect of pre-training, an additional condition was tested in which all weights were fully randomized and models were trained end-to-end. The evaluation results across these fine-tuning levels are summarized in Fig. \ref{fig:finetuning}, where the horizontal axes represent the degree of fine-tuning and the vertical axes show the metrics and their values. For each condition, 5-fold cross-validation was again used to obtain robust performance estimates. Additionally, three levels of L2 regularization (with $\lambda$=0, 1e-4, 1e-3) were applied to assess its potential influence on the observed trends.

According to Fig. \ref{fig:finetuning}, the first key observation is that when no fine-tuning is applied (i.e., left-most on the horizontal axes), models are skeptical with high precision, low recall,  and low self-confidence. While increasing regularization leads to modest improvements, model self-confidence rises significantly only after unfreezing deeper encoder blocks. Second, all metrics roughly plateau once block 5 is unfrozen, with Fig. \citesubfig{fig:finetuning}{e} showing the highest detected defect count when $\lambda$=1e-3. Third, when trained from scratch (i.e., right-most on the horizontal axes), both certainty and abundance drop substantially. Furthermore, detected defect count decreases by 25\% for $\lambda$=1e-3. These results indicate pre-training is necessary for maximizing performance with EfficientNet encoders, and that fine-tuning deeper layers up to block 5 leads to maximum self-confidence.

The trends seen in Fig. \ref{fig:finetuning} provide some insight into the abilities of EfficientNet to adapt its pre-trained weights for a new task and dataset. The steep rise in F1-score when the first few blocks (i.e., blocks 1-2) become trainable implies that the most important dissimilarity between ImageNet and the TEM dataset used in this work is the fine image structures. That is, most improvement in F1-score comes from developing shallow layers responsible for extracting localized features. However, developing deep layers up to block 5, which extract more global features, are crucial for building model self-confidence. This is most likely due to TEM images being much more homogeneous than natural scene images in ImageNet, and thus unique  weights for global feature identification must be learned \cite{Stuckner2022}. However, very abstract and compressed features discovered in the deepest layers remain relatively consistent between ImageNet and TEM datasets.
	\section{Conclusion}
\label{sec:conclusion}

In this work, the performance of U-Net on a TEM image dataset of nanocrystalline grain morphologies has been improved by using a pre-trained EfficientNetB7 encoder with deep fine-tuning and L2-regularization. Two novel metrics, prediction certainty and abundance, have been introduced to provide a qualitative evaluation independent of a ground truth. A third metric, detected defect count, was also used to provide a practical and holistic evaluation of model performance. 

Replacing the encoder with EfficientNetB7 gave models excess capacity and redundancy, and by using L2-regularization, they obtained substantially higher prediction confidence. In total, 64\% more grains were detected over the unregularized baseline U-Net architecture. To explain these benefits, it was proposed that regularization enhances model self-confidence by preventing overfitting of arbitrary human errors, but is only useful when a large pre-trained encoder is used.

A deeper dive into the importance of encoder pre-training was conducted as well, where it was found that model self-confidence is only acquired by fine-tuning deep layers which have been pre-trained on ImageNet, perhaps due to the dissimilarity between TEM and natural scene images at each spatial level during encoding. By fine-tuning blocks 1-5, responsible for extracting local and global features, model performance reaches its maximum.
	\section*{Code and Data Availability}
\label{sec:code}

The code developed for this work can be found on GitHub at \href{https://github.com/psu-rdmap/unet-compare}{https://github.com/psu-rdmap/unet-compare}. Training results, accompanying analysis code, and the supplementary materials can be found on ScholarSphere at \href{https://scholarsphere.psu.edu/resources/b80356d7-6485-40aa-841d-8f598c4ee9e2}{https://scholarsphere.psu.edu/resources/b80356d7-6485-40aa-841d-8f598c4ee9e2}.

	\section*{Acknowledgments}
\label{sec:acknowledgments}

This work was supported by the Office of Nuclear Energy of U.S. Department of Energy under Contract No.DE-NE0009426. We thank Dr. Zefeng Yu and Prof. Arthur Motta of Pennsylvania State University for providing the original TEM images.

	{\small
		\bibliographystyle{ieeenat_fullname}
		\bibliography{11_references}

\begin{thebibliography}{23}
\providecommand{\natexlab}[1]{#1}
\providecommand{\url}[1]{\texttt{#1}}
\expandafter\ifx\csname urlstyle\endcsname\relax
  \providecommand{\doi}[1]{doi: #1}\else
  \providecommand{\doi}{doi: \begingroup \urlstyle{rm}\Url}\fi

\bibitem[Alrfou et~al.(2022)Alrfou, Kordijazi, and Zhao]{Alrfou2022}
Khaled Alrfou, Amir Kordijazi, and Tian Zhao.
\newblock Computer vision methods for the microstructural analysis of
  materials: The state-of-the-art and future perspectives.
\newblock 2022.

\bibitem[Alsabhan et~al.(2022)Alsabhan, Alotaiby, and Dudin]{Alsabhan2022}
Waleed Alsabhan, Turky Alotaiby, and Basil Dudin.
\newblock Detecting buildings and nonbuildings from satellite images using
  u-net.
\newblock \emph{Computational Intelligence and Neuroscience}, 2022:\penalty0
  1--13, 2022.

\bibitem[Bruno et~al.(2023)Bruno, Lynch, Jacobs, Morgan, and Field]{Bruno2023}
Gabriella Bruno, Matthew~J Lynch, Ryan Jacobs, Dane~D Morgan, and Kevin~G
  Field.
\newblock Evaluation of human-bias in labeling of ambiguous features in
  electron microscopy machine learning models.
\newblock \emph{Microscopy and Microanalysis}, 29:\penalty0 1493--1494, 2023.

\bibitem[Deng et~al.(2009)Deng, Dong, Socher, Li, Li, and Fei-Fei]{Deng2009}
Jia Deng, Wei Dong, Richard Socher, Li-Jia Li, Kai Li, and Li Fei-Fei.
\newblock Imagenet: A large-scale hierarchical image database.
\newblock In \emph{2009 IEEE Conference on Computer Vision and Pattern
  Recognition}, pages 248--255. IEEE, 2009.

\bibitem[Jacobs(2022)]{Jacobs2022}
Ryan Jacobs.
\newblock Deep learning object detection in materials science: Current state
  and future directions.
\newblock \emph{Computational Materials Science}, 211:\penalty0 111527, 2022.

\bibitem[Liu et~al.(2019)Liu, Cao, Wang, and Wang]{Liu2019}
Zhenqing Liu, Yiwen Cao, Yize Wang, and Wei Wang.
\newblock Computer vision-based concrete crack detection using u-net fully
  convolutional networks.
\newblock \emph{Automation in Construction}, 104:\penalty0 129--139, 2019.

\bibitem[McKinney et~al.(2020)McKinney, Sieniek, Godbole, Godwin, Antropova,
  Ashrafian, Back, Chesus, Corrado, Darzi, Etemadi, Garcia-Vicente, Gilbert,
  Halling-Brown, Hassabis, Jansen, Karthikesalingam, Kelly, King, Ledsam,
  Melnick, Mostofi, Peng, Reicher, Romera-Paredes, Sidebottom, Suleyman, Tse,
  Young, Fauw, and Shetty]{McKinney2020}
Scott~Mayer McKinney, Marcin Sieniek, Varun Godbole, Jonathan Godwin, Natasha
  Antropova, Hutan Ashrafian, Trevor Back, Mary Chesus, Greg~S. Corrado, Ara
  Darzi, Mozziyar Etemadi, Florencia Garcia-Vicente, Fiona~J. Gilbert, Mark
  Halling-Brown, Demis Hassabis, Sunny Jansen, Alan Karthikesalingam,
  Christopher~J. Kelly, Dominic King, Joseph~R. Ledsam, David Melnick, Hormuz
  Mostofi, Lily Peng, Joshua~Jay Reicher, Bernardino Romera-Paredes, Richard
  Sidebottom, Mustafa Suleyman, Daniel Tse, Kenneth~C. Young, Jeffrey~De Fauw,
  and Shravya Shetty.
\newblock International evaluation of an ai system for breast cancer screening.
\newblock \emph{Nature}, 577:\penalty0 89--94, 2020.

\bibitem[Ophus(2019)]{Ophus2019}
Colin Ophus.
\newblock Four-dimensional scanning transmission electron microscopy (4d-stem):
  From scanning nanodiffraction to ptychography and beyond.
\newblock \emph{Microscopy and Microanalysis}, 25:\penalty0 563--582, 2019.

\bibitem[Ronneberger et~al.(2015)Ronneberger, Fischer, and
  Brox]{Ronneberger2015}
Olaf Ronneberger, Philipp Fischer, and Thomas Brox.
\newblock U-net: Convolutional networks for biomedical image segmentation.
\newblock 2015.

\bibitem[Shen et~al.(2021{\natexlab{a}})Shen, Li, Wu, Liu, Greaves, Hao,
  Krakauer, Krudy, Perez, Sreenivasan, Sanchez, Torres-Velázquez, Li, Field,
  and Morgan]{Shen2021_2}
Mingren Shen, Guanzhao Li, Dongxia Wu, Yuhan Liu, Jacob~R.C. Greaves, Wei Hao,
  Nathaniel~J. Krakauer, Leah Krudy, Jacob Perez, Varun Sreenivasan, Bryan
  Sanchez, Oigimer Torres-Velázquez, Wei Li, Kevin~G. Field, and Dane Morgan.
\newblock Multi defect detection and analysis of electron microscopy images
  with deep learning.
\newblock \emph{Computational Materials Science}, 199:\penalty0 110576,
  2021{\natexlab{a}}.

\bibitem[Shen et~al.(2021{\natexlab{b}})Shen, Li, Wu, Yaguchi, Haley, Field,
  and Morgan]{Shen2021_1}
Mingren Shen, Guanzhao Li, Dongxia Wu, Yudai Yaguchi, Jack~C. Haley, Kevin~G.
  Field, and Dane Morgan.
\newblock A deep learning based automatic defect analysis framework for in-situ
  tem ion irradiations.
\newblock \emph{Computational Materials Science}, 197:\penalty0 110560,
  2021{\natexlab{b}}.

\bibitem[Siddique et~al.(2021)Siddique, Paheding, Elkin, and
  Devabhaktuni]{Siddique2021}
Nahian Siddique, Sidike Paheding, Colin~P. Elkin, and Vijay Devabhaktuni.
\newblock U-net and its variants for medical image segmentation: A review of
  theory and applications.
\newblock \emph{IEEE Access}, 9:\penalty0 82031--82057, 2021.

\bibitem[Stuckner et~al.(2022)Stuckner, Harder, and Smith]{Stuckner2022}
Joshua Stuckner, Bryan Harder, and Timothy~M. Smith.
\newblock Microstructure segmentation with deep learning encoders pre-trained
  on a large microscopy dataset.
\newblock \emph{npj Computational Materials}, 8:\penalty0 200, 2022.

\bibitem[Tan and Le(2019)]{Tan2019}
Mingxing Tan and Quoc~V. Le.
\newblock Efficientnet: Rethinking model scaling for convolutional neural
  networks.
\newblock 2019.

\bibitem[Wagner et~al.(2019)Wagner, Sanchez, Tarabalka, Lotte, Ferreira, Aidar,
  Gloor, Phillips, and Aragão]{Wagner2019}
Fabien~H. Wagner, Alber Sanchez, Yuliya Tarabalka, Rodolfo~G. Lotte, Matheus~P.
  Ferreira, Marcos P.~M. Aidar, Emanuel Gloor, Oliver~L. Phillips, and Luiz E.
  O.~C. Aragão.
\newblock Using the u‐net convolutional network to map forest types and
  disturbance in the atlantic rainforest with very high resolution images.
\newblock \emph{Remote Sensing in Ecology and Conservation}, 5:\penalty0
  360--375, 2019.

\bibitem[Wang et~al.(2021)Wang, Du, Wu, Li, Zhao, and Zhu]{Wang2021}
Chunshan Wang, Pengfei Du, Huarui Wu, Jiuxi Li, Chunjiang Zhao, and Huaji Zhu.
\newblock A cucumber leaf disease severity classification method based on the
  fusion of deeplabv3+ and u-net.
\newblock \emph{Computers and Electronics in Agriculture}, 189:\penalty0
  106373, 2021.

\bibitem[Wang et~al.(2022)Wang, Jin, Wong, Chen, Bei, Wang, Ziatdinov, Weber,
  Zhang, Poplawsky, and More]{Wang2022}
Xing Wang, Ke Jin, Chun~Yin Wong, Di Chen, Hongbin Bei, Yongqiang Wang, Maxim
  Ziatdinov, William~J. Weber, Yanwen Zhang, Jonathan Poplawsky, and Karren~L.
  More.
\newblock Understanding effects of chemical complexity on helium bubble
  formation in ni-based concentrated solid solution alloys based on elemental
  segregation measurements.
\newblock \emph{Journal of Nuclear Materials}, 569:\penalty0 153902, 2022.

\bibitem[Williams and Carter(2009)]{Williams2009}
David~B. Williams and C.~Barry Carter.
\newblock \emph{Transmission Electron Microscopy}.
\newblock Springer US, 2009.

\bibitem[Xu et~al.(2024)Xu, Yu, Chen, Chen, Motta, and Wang]{Xu2024}
Xinyuan Xu, Zefeng Yu, Wei-Ying Chen, Aiping Chen, Arthur Motta, and Xing Wang.
\newblock Automated analysis of grain morphology in tem images using
  convolutional neural network with chac algorithm.
\newblock \emph{Journal of Nuclear Materials}, 588:\penalty0 154813, 2024.

\bibitem[Yu et~al.(2022)Yu, Xu, Chen, Sharma, Wang, Chen, Ulmer, and
  Motta]{Yu2022}
Zefeng Yu, Xinyuan Xu, Wei-Ying Chen, Yogesh Sharma, Xing Wang, Aiping Chen,
  Christopher~J. Ulmer, and Arthur~T. Motta.
\newblock Citation hidden due to anonymity.
\newblock \emph{Acta Materialia}, 231:\penalty0 117856, 2022.

\bibitem[Zhou et~al.(2023)Zhou, Huang, Pu, Guan, Huang, and Ling]{Zhou2023}
Caixia Zhou, Yaping Huang, Mengyang Pu, Qingji Guan, Li Huang, and Haibin Ling.
\newblock The treasure beneath multiple annotations: An uncertainty-aware edge
  detector.
\newblock 2023.

\bibitem[Zhou et~al.(2024)Zhou, Huang, Pu, Guan, Deng, and Ling]{Zhou2024}
Caixia Zhou, Yaping Huang, Mengyang Pu, Qingji Guan, Ruoxi Deng, and Haibin
  Ling.
\newblock Muge: Multiple granularity edge detection.
\newblock In \emph{2024 IEEE/CVF Conference on Computer Vision and Pattern
  Recognition (CVPR)}, pages 25952--25962. IEEE, 2024.

\bibitem[Zhou et~al.(2018)Zhou, Siddiquee, Tajbakhsh, and Liang]{Zhou2018}
Zongwei Zhou, Md~Mahfuzur~Rahman Siddiquee, Nima Tajbakhsh, and Jianming Liang.
\newblock Unet++: A nested u-net architecture for medical image segmentation.
\newblock 2018.

\end{thebibliography}
	}

\end{document}